\def\es{\epsilon_s}
\def\os{\Omega_s}

\def\estar{\epsilon^{\ast}}
\def\edstar{\epsilon_D^{\ast}}

\def\ostar{\Omega^{\ast}}
\def\odstar{\Omega_D^{\ast}}
\def\mudstar{\mu_D^{\ast}}

\def\mustar{\mu^{\ast}}
\def\bg{\beta_{\Gamma}}

\documentstyle[11pt, aaspp4, epsf, rotate]{article}
\begin{document}

\title{Analyzing the Multiwavelength Spectrum of BL Lacertae During 
the July 1997 Outburst}

\author{M. B\"ottcher\footnote{Chandra Fellow}}
\affil{Department of Space Physics and Astronomy, Rice University, MS 108, \\
6100 S. Man Street, Houston, TX 77005 - 1892}
\author{S. D. Bloom {\altaffilmark{2,}\altaffilmark{3}}}
\affil{Laboratory for High Energy Astrophysics, NASA/Goddard Space
Flight Center, \\
Greenbelt, MD 20771}

\altaffiltext{2}{NAS/NRC Resident Research Associate}
\altaffiltext{3}{Current Address: Infrared Processing and Analysis
Center, Jet Propulsion Laboratory and California Institute of Technology,
MS 100-22, Pasadena, CA 91125}

\vskip 4mm

\centerline{\sl To appear in The Astronomical Journal}

\begin{abstract}
The multiwavelength spectrum of BL~Lacertae during its July 1997
outburst is analyzed in terms of different variations of the
homogeneous leptonic jet model for the production of high-energy 
radiation from blazars. We find that a two-component $\gamma$-ray 
spectrum, consisting of a synchrotron self-Compton and an external 
Compton component, is required in order to yield an acceptable 
fit to the broadband spectrum. Our analysis indicates that in 
BL~Lac, unlike other BL~Lac objects, the broad emission line 
region plays an important role for the high-energy emission. 
Several alternative blazar jet models are briefly discussed.
In the appendix, we describe the formalism in which the
process of Comptonization of reprocessed accretion disk
photons is treated in the previously developed blazar jet 
simulation code which we use.
\end{abstract}


\section{Introduction}
A dramatic gamma-ray outburst of BL Lacertae (z=0.069) during 
July 1997 was recently reported by \cite{blo97}. BL Lac's gamma-ray 
activity was coincident with bright and variable emission from optical 
to X-rays (\cite{nob97}; \cite{mat96}); however, the light curves in 
each waveband are too sparsely sampled to determine accurately whether 
they are correlated. In addition, the gamma-ray spectrum was 
significantly harder than in the lower intensity state, which is 
commonly seen for flaring blazars (\cite{muk96}; \cite{muk97}). 
Multiwavelength flares coincident with gamma-ray flares have also 
been documented for many other objects (eg., 3C 279, \cite{weh98}; 
and 1406-076, \cite{wag95}). The observations during the flare 
will be discussed further in Section 2.

Historically, BL Lacertae was the prototypical `BL Lac Object', 
i.e., has quasar-like continuum properties, but does usually not 
show emission or absorption lines (the equivalent width is $<5 $ 
\AA). Though, H$\alpha$ emission lines have been detected in 
BL Lac itself during a period of several weeks in 1995 (\cite{ver95}; 
\cite{cor96}). BL Lacertae is also a member of the blazar class of 
objects, i.e. a radio-loud AGN which has bright and variable 
broadband continuum emission as well as relatively high polarization. 
About 50 blazars have been detected in high energy gamma-rays by the 
EGRET instrument, and 14 are BL~Lac objects (\cite{muk97}).

In this paper we will examine the multiwavelength spectral energy
distribution (SED; or plot of $\nu F_\nu$) of BL Lac considering 
several emission mechanisms as the possible source of gamma-rays: 
synchrotron self-Compton scattering (SSC; \cite{ghi85}; \cite{blo96}), 
external inverse Compton scattering with an accretion disk as the 
source of soft photons (ECD; \cite{der92}), and an external inverse
Compton scattering of soft photons reprocessed in broad line region 
clouds (ECC; \cite{sik94}). These mechanisms and the techniques 
used will be discussed further in Section 3. The modeling results 
and their implications for the nuclear environment of BL~Lac are 
presented and discussed in Section 4. A discussion of the application
of alternative --- leptonic as well as hadronic --- models to the
broadband spectrum on BL~Lacertae follows in Section 5. We summarize 
in Section 6. In the appendix, we discuss our numerical treatment of 
Compton scattering of reprocessed accretion disk radiation, which 
was not included in the original version of the jet simulation 
code used for our modeling procedure.

\section{Observations}

In addition to the optical ($V$ and $R$ band) and EGRET gamma-ray 
observations of Bloom {\it et al.} (\markcite{blo97}1997), we have 
included additional optical ($BVRI$; \cite{web98}), radio, X-ray, 
and low energy gamma-ray observations from the time of the outburst. 
The radio observations at 2.25 and 8.3 GHz are from the Green Bank 
Interferometer monitoring program (\cite{wal96}), the 4, 8 and 14~GHz 
measurements are from the UMRAO monitoring program  (\cite{all85}). 
The 2-10 keV ASCA X-ray spectrum (\cite{mak97}) shown in Figs. 1 -- 3 
(thick dashed curves) are from continuous measurements of July
18.6 -- 19.62 and yielded a photon spectral index of $1.44 \pm 0.01$.
The ASCA spectrum agrees well with the XTE PCA spectrum of Madejski 
et al. (\markcite{mad97a}1997a) from the same epoch, which is not
plotted in the figures. Madejski et al. (\markcite{mad99}1999) show
that the X-ray spectrum is harder than for earlier epochs, and that
it is variable on time scales of approximately one day. The OSSE 
observations from 0.15 -- 1.0 MeV have been provided by the OSSE 
Team (\cite{gro97}). 

TeV upper limits from CAT (\cite{fle97}) and HEGRA (\cite{aha99}) 
from observations in 1997 August are also added (filled square and 
circle, respectively). Although these measurements are not quite 
simultaneous to the remainder of the observations, they might give 
a rough estimate of the actual TeV flux at the time of the July 
1997 outburst because the optical flux during the TeV observations 
was similar to the flux observed during the EGRET flare.

A reddening correction of $A_{\rm R}=0.7$ derived from $E(B-V) = 0.3$
from Vermeulen {\it et al.} (\markcite{ver95}1995) and the extinction 
law of Cardelli {\it et al.} (\markcite{car89}1989) were applied to 
the optical data presented in Bloom {\it et al.} (\markcite{blo97}1997).
Webb {\it et al.} (\markcite{web98}1998) have used a similar correction 
($E(B-V)=0.36$). The values of the optical flux plotted in Figs. 1--3 
are averages during the week of the gamma-ray flare. The optical flux 
exhibited short-term variability on a timescale of several hours. The 
error bars shown in the figures correspond to the dispersion of the
optical fluxes during the observing period.

\section{Outline of the model}

We modelled the broadband spectrum of BL~Lac with the pair plasma 
jet simulation code described in detail in B\"ottcher {\it et al.}
(\markcite{bms97}1997, hereafter BMS). In this section, we only 
give a short outline of the model and the assumptions.

Our model is based on various versions of leptonic jet models
for blazar emission. A blob of ultrarelativistic pair plasma is 
moving outward from the central accretion disc along a pre-existing 
cylindrical jet structure, with relativistic speed $\bg c$ and bulk 
Lorentz factor $\Gamma$. In the following, primed quantities refer
to the rest frame comoving with the relativistic plasma blob. At 
the time of injection into the jet at height $z_i$ above the 
accretion disc, the pair plasma is assumed to have an isotropic 
momentum distribution in the comoving frame. The distribution 
functions at this time are power-laws, $n'(\gamma') = n_0 \, 
{\gamma'}^{-s}$, with low- and high-energy cut-offs $\gamma_1 
\le \gamma' \le \gamma_2$. The total density of electrons (and 
positrons) in the comoving frame is $n'_e$. The blob is spherical 
in the comoving frame with radius $R'_B$ which does not change 
along the jet. A randomly oriented magnetic field of strength 
$B' \lesssim B'_{ep}$ is present, where $B'_{ep}$ is the magnetic 
field corresponding to equipartition with the energy density 
of pairs at the base of the jet. The jet is inclined at an 
angle $\theta_{obs}$ with respect to the line of sight. The 
spherical geometry in the comoving rest frame implies that,
in the stationary frame of the AGN, the blob appears Lorentz
contracted with a length parallel to the jet axis of $R_{\parallel}
= R'_B / \Gamma$ and a perpendicular extent $R_{\perp} =
R'_B$.

As the blob moves out, various radiation and cooling mechanisms
are at work: synchrotron radiation, synchrotron self-Compton
radiation (SSC; \cite{ghi85}, \cite{mg85}, \cite{mgc92}, 
\cite{blo96}), inverse-Compton scattering of external 
radiation from the accretion disc, either entering the jet 
directly (ECD for External Compton scattering of direct Disc
radiation; \cite{der92}, \cite{der93}) or after being
rescattered by surrounding (BLR) material (ECC for External
Compton scattering of radiation from Clouds; \cite{sik94}, 
\cite{bl95}, \cite{dss97}). The latter process was not 
included in the original version of the BMS simulation 
code. Therefore, its treatment is described in more
detail in the appendix.

The BLR is represented by a spherical layer of material with uniform 
density and total Thomson depth $\tau_{T,BLR}$, extending between the 
radii $r_{in, BLR} \le r \le r_{out, BLR}$ from the central black hole. 
The synchrotron mirror mechanism (\cite{gm96}, \cite{bed98}) 
is unlikely to be efficient in BL~Lac objects (\cite{bd98}), 
because it requires a high Thomson depth of the surrounding
material ($\tau_{T,BLR} \gtrsim 0.1$), for which no evidence
exists in this class of objects. Therefore we do not consider 
it in our spectral modeling procedure for BL~Lac.

The central accretion disc is radiating a standard Shakura-Syunyaev
(\markcite{ss73}1973) disc spectrum around a black hole of
$10^6 \, M_6$ solar masses. The total disc luminosity is 
$10^{44} \, L_{44}$~erg~s$^{-1}$. The full angle dependence
of the disc radiation field, according to its radial temperature
structure, is taken into account. Compton scattering of all
radiation fields is calculated using the full Klein-Nishina 
cross section. The SSC mechanism is calculated to arbitrarily 
high scattering order.

After the initial injection of ultrarelativistic pairs, all of
the mechanisms mentioned above (except the synchrotron mirror)
are taken into account in order to follow the evolution of the
electron/positron distribution functions and the time-dependent
broadband emission as the blob moves out along the jet. 
$\gamma\gamma$ absorption intrinsic to the source and the
re-injection of pairs due to $\gamma\gamma$ pair production is
taken into account completely self-consistently using the exact,
analytic solution of B\"ottcher \& Schlickeiser (\markcite{bs97}1997) 
for the pair injection spectrum due to $\gamma\gamma$ absorption. 
However, with the parameters adopted to fit the spectrum of BL~Lac, 
the blob is optically thin to $\gamma\gamma$ absorption for photons
up to at least 100~GeV so that this process plays a minor role.
Consequently, since we assume that the blob does not expand, the 
total electron (positron) density $n'_e$ in the blob remains 
basically constant along the jet.

The broadband spectrum investigated in this paper is a time 
average over the entire duration of the $\gamma$-ray flare. 
Thus, it corresponds to the time average of the radiation emitted 
by one or multiple plasma blobs rather than to a snapshot during 
the evolution of a single blob. To compare our model spectra 
to the observed broadband spectrum of BL~Lacertae, we use the 
time-integrated broadband emission (fluence) of a single blob. An 
average flux is calculated by dividing the fluence by a representative 
time scale which corresponds to the time scale of repeated blob ejection 
events. A jet quasi-continuously filled with relativistic pair plasma 
corresponds to a repetition time of $t_{rep} = 2 R'_B / (\Gamma \, c)$. 
We define a filling factor $f$ of the jet given by $f = t_{rep}^{cont} 
/ t_{rep}$. In all our model fits we find that the repetition time 
scale (in the comoving frame) is generally longer than the cooling 
time scale for the highest-energy electrons, but comparable to 
the decay time scale of radiation produced by the inefficiently 
cooled part of the electron spectrum. Therefore, the variability 
in the high-energy portions of both the synchrotron and the 
inverse-Compton components (optical --- UV and $> 100$~MeV 
$\gamma$-rays, respectively) is dominated by light-travel time
effects within the source. These effects have recently been 
studied in detail by Chiaberge \& Ghisellini (\markcite{chia98}1998). 
Strong variability on the light-travel time scale through the blob 
is expected in the high-energy parts of both radiation components, 
if the filling factor is substantially less than 1. The variability 
in these two energy bands (optical/UV and high-energy $\gamma$-rays) 
is expected to be correlated. If $f \sim 1$, any variability pattern 
is determined by variations of the electron injection parameters 
(i. e. the electron density, cutoff energies, spectral index etc.) 
and its modelling would require additional model assumptions 
about the underlying accretion flow and the pair injection
mechanism. The low-frequency portions of both components 
(radio --- infrared and X-rays, respectively), which are 
predominantly produced by basically uncooled electrons, 
are expected to show relatively weak variability on longer 
time scales, also determined by long-term variations of 
the pair injection parameters.

We point out that the evolution of individual blobs is virtually
independent of the existence of multiple blobs along the jet if
there are no significant fluctuations of the bulk Lorentz factor 
$\Gamma$ within the flow. The radiation emitted by each individual 
blob is quasi-isotropic in the comoving frame so that the radiative 
interaction between different blobs may be neglected. The situation
investigated here is geometrically very similar to shocks propagating
along the jet because for large values of $\Gamma$ the blobs are
strongly Lorentz contracted along the jet axis. A major difference
between a shock-in-jet model and our impulsive-acceleration model
is that in the case of shocks particle acceleration will continue
to take place at large distances from the central engine, where
the external radiation field due to direct or reprocessed accretion
disk radiation is negligible. Therefore, those models will be strongly
SSC dominated. The spectral signatures will be basically indistinguishable
from the SSC dominated cases of our model calculations, which we will 
discuss in the following section.

Throughout this paper, we assume $H_0 = 65$~km~s$^{-1}$~Mpc$^{-1}$
and $q_0 = 0.5$.

\section{Modeling results and discussion}

Apart from the observed broadband spectrum, there are several
other observational results on BL~Lac, which constrain the
choice of parameters to be adopted in our modeling efforts.
Recently, Denn et al. (\markcite{denn99}1999) have observed 
superluminal motion in BL~Lac of single components with up to
$\beta_{\perp, app.} = (5.0 \pm 0.2) \, h^{-1} = 7.7 \pm 0.3$. 
This indicates that the typical bulk Lorentz factor is 
expected to be in the range $\Gamma \gtrsim 8$. This is much
higher than the limit implied by the required transperency of
the emitting region for high-energy $\gamma$-ray photons due
to $\gamma\gamma$ absorption within the source, which yields

\begin{equation}
D \gtrsim \left( {L_{\gamma} \, \sigma_T \, (1 + z) \over 
4 \pi \, m_e c^4 \, \Delta t_{var}} \right)^{1/5} \approx
1.4
\end{equation}
for a variability time scale of 1 day. Here $D = \left(\Gamma \,
[1 - \beta_{\Gamma} \, \cos\theta_{obs}]\right)^{-1}$ is the 
Doppler factor.

Occasionally, BL~Lac shows strong emission lines. On May 21 and
June 1, 1995, their width has been measured to be $\langle v 
\rangle \approx 4000$~km~s$^{-1}$ (\cite{ver95}, \cite{cor96}). 
If we attribute this width to Keplerian motion, we find for the 
average distance of the BLR clouds from the central black hole:

\begin{equation}
\overline r_{BLR} = 2.77 \cdot 10^{-4} \, M_6 \, {\rm pc}.
\end{equation}
We have used the XSTAR code (Kallman \& Krolik \markcite{kk98}1998)
in order to estimate the emission line luminosity resulting from
a BLR consisting of Thomson thick ($\tau_T \sim 10$) individual 
clouds, illuminated by disk blackbody radiation from an accretion 
disc around a black hole of $M_6 \approx 2$, radiating at a luminosity 
of $L_D \sim 10^{44}$~erg~s$^{-1}$. We find that an angle-average 
Thomson depth $\tau_{T,BLR} \sim 0.02$ is sufficient to produce the 
observed line luminosity $L_{H\alpha} \sim 10^{41}$~erg~s$^{-1}$. 
However, this value of the required $\tau_{T, BLR}$ is strongly 
dependent on the clumpiness of the BLR and may well deviate by 
more than an order of magnitude from the value given above.

The total luminosity of the accretion disc is poorly constrained
since the spectral coverage is insufficient to clearly exclude
the existence of a big blue bump from the accretion disc. We
adopt a disc luminosity of $L_D = 10^{44}$~erg~s$^{-1}$, implying
that no big blue bump should be observable above the synchrotron
spectrum.

The shortest observed variability timescale during the 1997 
flare was several hours, observed at optical wavelengths. The 
variability timescale at $\gamma$-ray energies was $\sim 1$~day,
while it was slightly shorter than a day in X-rays. The optical 
variability yields a constraint on the extent of the emission 
region of 

\begin{equation}
R'_B \lesssim 5.4 \cdot 10^{14} \, D \, \Delta t_5 \, {\rm cm}
\label{RB}
\end{equation}
where $\Delta t_5$ is the variability timescale in units of 
5~hours.

We have first attempted to fit the broadband spectrum of BL~Lac 
with the basic models, where one of the radiation mechanisms
mentioned above dominates the entire X-ray and $\gamma$-ray 
spectrum.

Madejski et al. (\markcite{mad99}1999) have presented an analytical 
estimate of the required parameters for an SSC fit to BL~Lacertae, 
indicating that, if the entire $\gamma$-ray spectrum, peaking at 
$\epsilon_{SSC} \gtrsim 2 \cdot 10^4$, where $\epsilon = h \nu 
/ (m_e c^2)$ is the dimensionless photon energy, were produced by 
the SSC mechanism, unrealistic values of $\Gamma \gtrsim 100$ 
and $B' \lesssim 10^{-4}$~G would be needed. As mentioned in
the previous section, the same arguments would apply to an SSC
dominated shock-in-jet model.

\begin{figure}
\epsfysize=12cm
\rotate[r]{
\epsffile[150 0 580 550]{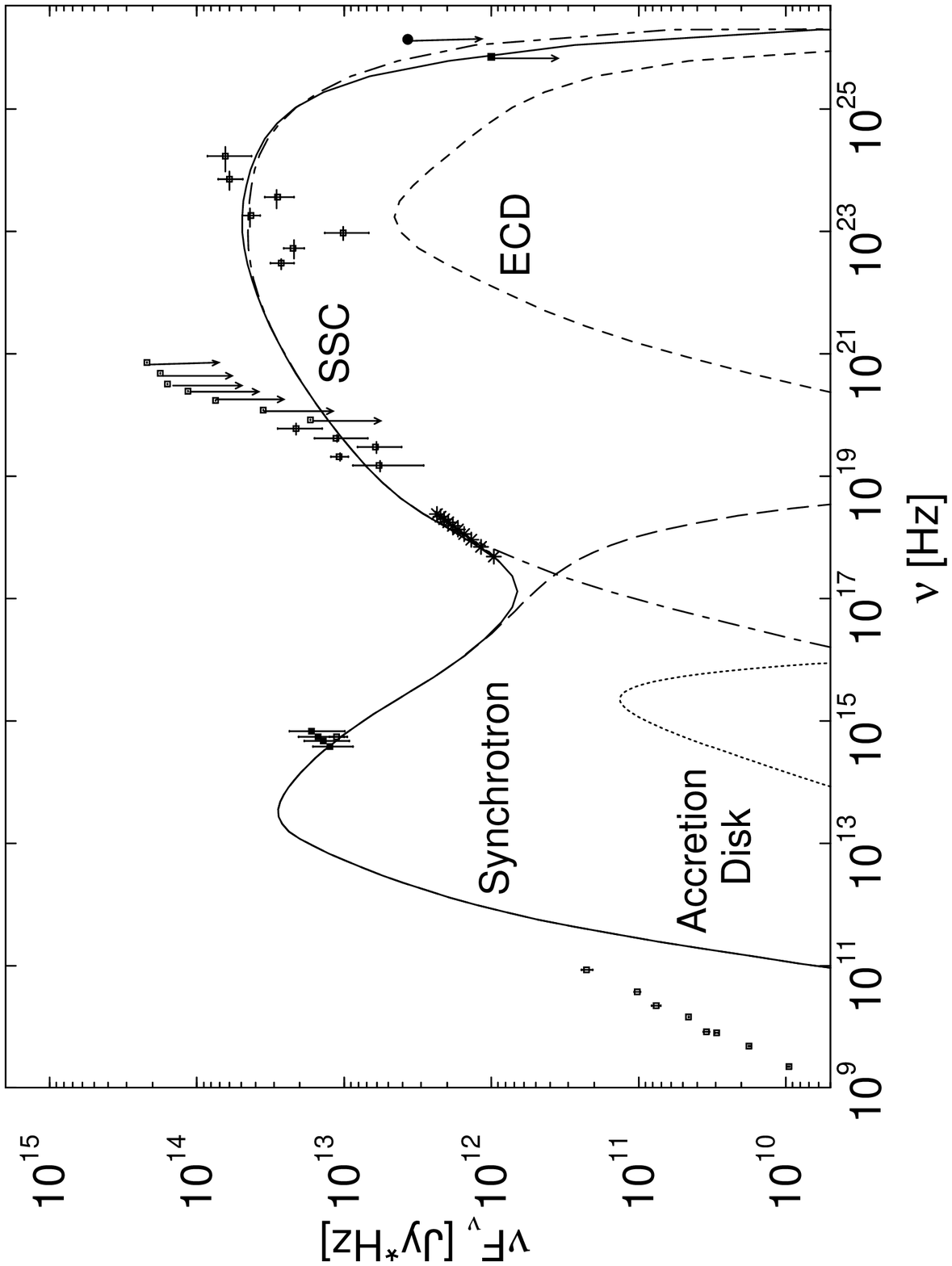}}
\caption[]{Attempt to fit the SED of BL~Lac with an SSC-dominated model.
Parameters: $\gamma_1 = 3000$, $\gamma_2 = 10^5$, $s = 2.6$, $B' = 1.2$~G, 
$\Gamma = 15$, $\theta_{obs} = 2^o$, $n'_e = 10^3$~cm$^{-3}$,
$R'_B = 10^{15}$~cm, $z_i = 10^{-3}$~pc, $f = 0.6$}
\label{ssc_fit}
\end{figure}

Fig. \ref{ssc_fit} illustrates our attempts to fit a strongly 
SSC-dominated model to the broadband spectrum of BL~Lac. The optical 
flux and the hard X-ray and soft $\gamma$-ray spectrum are well 
reproduced by this model. It is also very well consistent with 
the concave spectral shape at soft X-rays for which Madejski et al. 
(\markcite{mad99}1999) have found weak evidence in ASCA observations
of BL~Lacertae in 1995 November, suggesting that the high-energy
tail of the synchrotron component and the low-energy tail of the
inverse-Compton component overlap in the 2 -- 10~keV energy band. 
However, a pure SSC model is not able to reproduce the hard 
power-law above 100~MeV (EGRET). This is primarily because an 
SSC spectrum does not produce a strong spectral break in the 
MeV range which would be necessary to connect the X-ray / soft 
$\gamma$-ray spectrum to the EGRET spectrum. Our model $\gamma$-ray
spectrum peaks at a values of $\epsilon_{ssc}$ significantly lower 
than $2 \cdot 10^4$, which is the only way to allow for more
realistic values of $\Gamma$ and $B'$ than estimated above.

\begin{figure}
\epsfysize=12cm
\rotate[r]{
\epsffile[150 0 580 550]{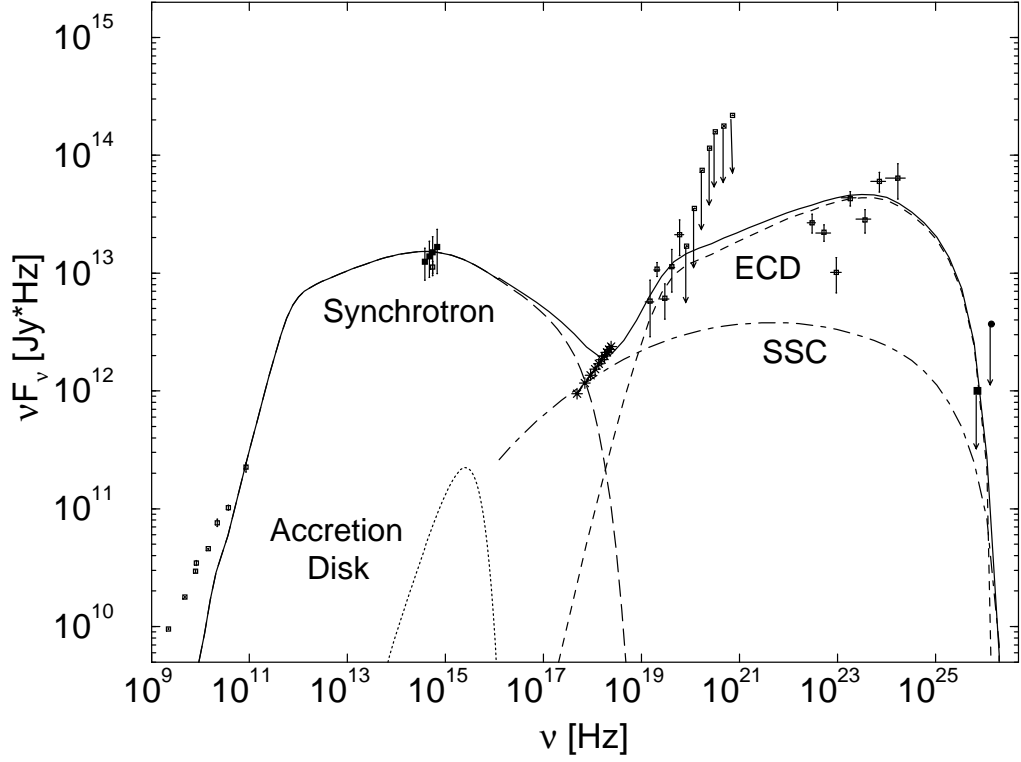}}
\caption[]{Attempt to fit the SED of BL~Lac with an ECD-dominated model.
Parameters: $\gamma_1 = 30$, $\gamma_2 = 10^5$, $s = 2.6$, $B' = 0.8$~G, 
$\Gamma = 20$, $\theta_{obs} = 2^o$, $n'_e = 400$~cm$^{-3}$,
$R'_B = 2 \cdot 10^{16}$~cm, $z_i = 7 \cdot 10^{-3}$~pc, $M_6 = 100$, 
$f = 1$}
\label{ecd_fit}
\end{figure}

\begin{figure}
\epsfysize=12cm
\rotate[r]{
\epsffile[150 0 580 550]{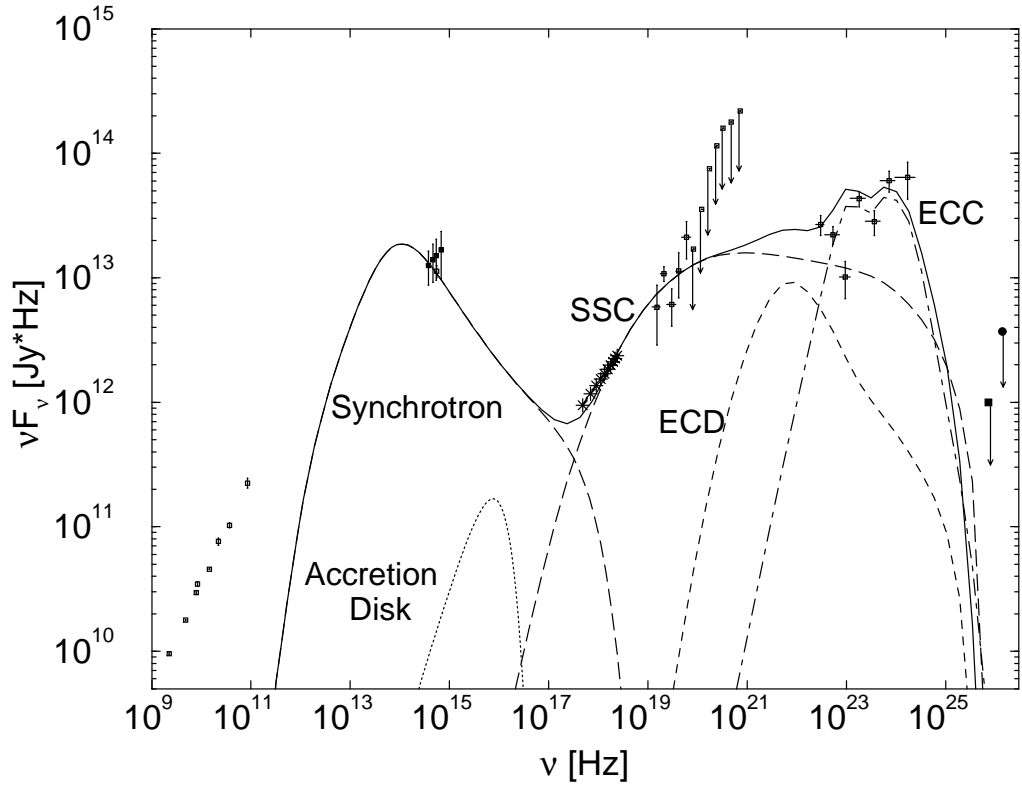}}
\caption[]{Fit to the SED of BL~Lac with a combined SSC + ECC model.
Parameters: $\gamma_1 = 500$, $\gamma_2 = 3.5 \cdot 10^4$, $s = 2.5$,
$B' = 9$~G, $\Gamma = 15$, $\theta_{obs} = 3^o$, $n'_e = 3 \cdot 
10^3$~cm$^{-3}$, $R'_B = 1.2 \cdot 10^{15}$~cm, $z_i = 5 \cdot 10^{-4}$~pc, 
$r_{in,BLR} = 5 \cdot 10^{-4}$~pc, $r_{out,BLR} = 6 \cdot 10^{-4}$~pc,
$\tau_{T,BLR} = 0.025$, $M_6 = 2$, $f = 0.8$}
\label{combined_fit}
\end{figure}

In contrast to the SSC model, a pure external inverse-Compton 
model does produce the necessary spectral break in the MeV
energy range and therefore reproduces the OSSE -- EGRET 
$\gamma$-ray spectrum of BL~lac better (see Fig. 2). However,
the spectrum predicted by the ECD-dominated model is inconsistent
with the ASCA X-ray spectrum. The jet filling factor corresponding 
to the simulation shown in Fig. \ref{ecd_fit} is $f \approx 1$, 
which implies that any variability would be related to variations 
of the particle injection/acceleration mechanism. In order to have 
the ECD component strongly dominate the SSC component we need 
to assume a much larger mass of the central black hole ($M_6 
= 100$) than for our best fit presented in Fig. \ref{combined_fit},
where $M_6 = 2$.

The parameters required for an synchrotron + ECD model to
reproduce the observed luminosities and peak frequencies
in the broadband spectrum of BL~Lacertae can be estimated
using

\begin{equation}
{L_{ECD} \over L_{sy}} \approx {u'_{ECD} \over u'_B} \approx
{2 \, L_D \over {B'}^2 z_i^2 \, c \, \Gamma^2}, 
\label{L_ECD}
\end{equation}
where $z_i$ is the injection height of the plasma blob above
the accretion disk, which is approximated by a point source 
located at the center of the AGN. Using

\begin{equation}
B' \approx B_{cr} \, {\epsilon_D \, \epsilon_{sy} \over
\Gamma \, \epsilon_{ECD}} \sim 0.9 \, \Gamma^{-1} \, {\rm G},
\end{equation}
where $B_{cr} = 4.414 \cdot 10^{13}$~G, we find that eq. 
\ref{L_ECD} becomes independent of $\Gamma$,
and

\begin{equation}
z_i \sim {\sqrt{2} \over \sqrt{c} \, B_{cr}} \, \sqrt{L_D \, L_{sy} 
\over L_{ECD}} \, {\epsilon_{ECD} \over \epsilon_D \, \epsilon_{sy}}
\sim 1.6 \cdot 10^{-2} \, {\rm pc},
\end{equation}
where $\epsilon_D \sim 10^{-4}$ is the dimensionless peak energy of 
the accretion disk spectrum.

Since in the ECC process the accretion disc photon energies 
are boosted to much higher energies into the blob rest frame, 
because most of the soft photons enter the blob from the front,
it is not possible to produce the OSSE -- EGRET spectrum of
BL~Lac with a single ECC radiation component. However, it is
appropriate to produce the observed $\gamma$ radiation in the
EGRET energy range and, in particular, the spectral peak at
$\epsilon_{ECC} \gtrsim 2 \cdot 10^4$ with reasonable parameters.
Approximating the external photon energy density due to reprocessed
accretion disk photons as 

\begin{equation}
u'_{ECC} \approx {L_D \, \tau_{T, BLR} \, \Gamma^2 \over
4 \pi \, {\overline r_{BLR}}^2 \, c},
\end{equation}
we find
\begin{equation}
{L_{ECC} \over L_{sy}} \approx {2 \, L_D \, \tau_{T, BLR}
\over {\overline r_{BLR}}^2 \, c \, B_{cr}^2} \, \left(
{\epsilon_{ECC} \over \epsilon_{sy} \, \epsilon_D} \right)^2
\end{equation}
or
\begin{equation}
{{\overline r_{BLR}}^2 \over \tau_{T, BLR}} \sim 2.3 \cdot 10^{33}
\> {\rm cm}^2,
\end{equation}
in reasonable agreement with our estimates on $\overline r_{BLR}$ and 
$\tau_{T, BLR}$ as given at the beginning of this section, considering 
the uncertainty of those estimates due to the unknown degree of
clumpiness of the BLR. The co-moving magnetic field may be estimated
as

\begin{equation}
B' \sim B_{cr} \, {\epsilon_{sy} \over \epsilon_{ECC}} \, \epsilon_D
\, \Gamma \sim 0.8 \, \Gamma \> {\rm G}.
\end{equation}

The problems with fitting a one-component X-ray / $\gamma$-ray 
spectrum to the SED of BL~Lac lead us to the conclusion that a 
second component is necessary to explain its peculiar high-energy
spectrum. This is in agreement with the general tendency among
different subclasses of blazars found by Ghisellini et al.
(\markcite{ghi98}1998) that the sequence HBL $\to$ LBL $\to$
FSRQ may be related to an increasing dominance of external
Compton scattering over synchrotron self-Compton scattering
in the $\gamma$-ray regime. Since the properties of BL~Lacertae
are found to be intermediate between classical BL~Lac objects
and quasars, such a combination of SSC and ERC mechanisms appears 
to be a natural explanation for its peculiar high-energy spectrum.
For the case of BL~Lacertae this conclusion was first pointed
out on the basis of analytical estimates by Madejski {\it et al.} 
(\markcite{mad97b}1997b, \markcite{mad99}1999). Since the SSC 
mechanism yields a reasonable fit to the optical to soft 
$\gamma$-ray spectrum, we started out from this model and 
adjusted the parameters in a way that a second component 
shows up in the EGRET energy range. The fact that there is 
clear evidence for the existence of BLR clouds rather close 
to the central black hole, and the analytical estimates 
regarding this process suggest that the ECC mechanism is
responsible for this additional high-energy component.

Fig. \ref{combined_fit} shows our best fit to the SED of 
BL~Lac. The broadband spectrum from optical to high-energy 
$\gamma$-rays is reasonably well reproduced by our 
model. However, as all of our model calculations, it
underpredicts the radio flux, because the low-energy
cutoff in the electron energy distribution leads to
a sharp cutoff in the synchrotron spectrum below
$\sim 10^{13}$~Hz. Since there is no clear evidence
for a correlation between radio and $\gamma$-ray flares
in blazars (\cite{mue97}), it seems plausible that most 
of the radio flux is produced in the outer jet regions,
where the bulk kinetic energy of the jet is dissipated 
in interactions with external material. This is supported
by the results of long-term radio, IR, optical, and X-ray 
monitoring of BL~Lacertae as reported by Bregman et al. 
(\markcite{bregman90}1990). They have found that radio
flux variations typically lack fluctuations at optical
and IR frequencies by $\sim 1$ - 4~years. Their result
that radio and IR flux variations are correlated with X-ray 
variations, but not with fluctuations of the optical flux,
is consistent with the IR and the X-ray flux being produced 
by strongly cooled electrons in the jet via synchrotron and 
SSC radiation, respectively. The radio, IR, and X-ray
portions of the photon spectrum reflect long-term variations 
of the physical conditions at the acceleration site. Short-term
variations of the injected electron spectrum are smoothed out
by the slow cooling process. In contrast, the optical flux, 
produced by electrons cooling on much shorter time scales, 
reflects variations of the physical conditions on time scales 
comparable to the synchrotron cooling time scale of 
ultrarelativistic electrons. Using similar arguments,
McHardy et al. (\markcite{mchardy99}1999) have recently
interpreted simultaneous IR and X-ray fluctuations of
3C~273 as evidence for the SSC process being responsible
for the X-ray emission of that object.

The fact that in Figs. \ref{ssc_fit} and \ref{combined_fit} 
the total spectrum (solid line) is below the (unabsorbed) SSC 
spectrum (long dashed), indicates that here $\gamma\gamma$ 
absorption becomes important. As outlined above, $\gamma\gamma$ 
pair production is included in our simulations self-consistently. 

The blob repetition time required by our fit implies a
filling factor of $f = 0.8$, which means that the jet
is filled essentially continuously. As explained in the
previous section, this implies that due to light-travel
time effects within the blob no strong variability on 
time scales shorter than the light-travel time through 
the blob are expected (\cite{chia98}), although the
cooling timescale of ultrarelativistic electrons in the
blob rest frame is only a few $100$~s. Source variability
is directly related to variations of the injection/acceleration
mechanism.

Due to the very time-consuming nature of our simulations,
we cannot perform a detailed variational analysis of all
parameters and give relative errors on their values. 
However, we point out that the relative strength of the 
different radiation components as well as their location 
in energy space constrain most of the parameters listed 
in the figure captions rather tightly. For example, an
increase in the black hole mass leads to a) the BLR
moving outward (Eq. [1]) and thus to a decrease of the 
ECC component, and b) the Schwarzschild radius and therefore 
the disc moving outward and becoming cooler, implying that 
the ECD component becomes stronger (because photons enter 
the blob under a more favorable angle), but emerges at 
different energies (combined effect of different disc
temperature and different beaming pattern of disc
radiation into the blob). Due to the colder disk, the 
ECC component will emerge at lower energies. Such changes
also imply a change in the cooling time scale as compared
to the required repetition time scale, determined by the
total luminosity of a single blob which has consequences
on the predicted variability pattern as mentioned in the
previous section. This may indicate that the self-consistent 
system we simulate is rather sensitive to changes of the 
parameters, although there might still be ambiguities.

Our results are sensitive to the assumed injection height 
$z_i$ of the blob. According to the physical picture that 
the blob is produced by an explosive event in a hot corona
above the accretion disk, we have generally chosen $z_i 
\approx R_B$, since there is no obvious mechanism how 
relativistic particles could be transported to significant 
distances above the disk without radiative losses.

We find the appropriate average Thomson depth of the BLR,
$\tau_{T,BLR} = 0.025$, to be comparable to what we expect
in other BL~Lac objects (using the estimates given in
\cite{wandel97}). However, since the BLR seems to be 
located unusually close to the accretion disc ($\overline r
= 5.5 \cdot 10^{-4} \, {\rm pc} \approx 3 \cdot 10^3 \, R_S$), 
it can still produce a significant flux in H$\alpha$ line emission
and contribute a significant soft photon density from rescattered 
accretion disc radiation. As pointed out in B\"ottcher \&
Dermer (\markcite{bd98}1998), a BLR of such low Thomson
depth, located so close to the central black hole is
extremely inefficient in terms of the synchrotron mirror
mechanism, which provides additional justification for our 
neglect of this process.

According to the results of Stecker \& de Jager 
(\markcite{sdj97}1997), $\gamma\gamma$ absorption by 
the intergalactic infrared background radiation is 
negligible for these photon energies at the redshift of 
BL~Lac (z = 0.069), so the cutoff has to be intrinsic to
the source. However, the CAT and HEGRA measurements have 
been performed in August 1997 and were therefore not 
simultaneous to the rest of the broadband spectrum. Since 
generally blazars show the strongest and most rapid variability 
at the highest photon energies, the $> 300$~GeV flux during the 
EGRET outburst could well have been higher than indicated by the 
upper limits in Figs. \ref{ssc_fit} -- \ref{combined_fit}.

Our model has generally problems with the hard optical
spectrum. Although the correct optical flux is predicted,
rapid cooling leads to a steep optical synchrotron spectrum
$F_{\nu} \propto \nu^{-p/2}$ for a cooling particle population 
injected with a power-law $n(\gamma) \propto \gamma^{-p}$.
The same problem has been encountered in other blazars, such 
as PKS~0528+134 (\cite{muk99}) where, in some observing
periods, the optical spectrum indicates an upturn compared
to the overall synchrotron spectrum. This upturn is clearly
inconsistent with the onset of the inverse-Compton component
as well as with a ``big blue bump'' due to direct accretion
disk radiation, which, in the case of PKS~0528+134, would 
imply an accretion disk luminosity of $\sim 10^{48}$~erg~s$^{-1}$.
Up to now there is no generally accepted solution to this 
problem, and in most modeling efforts on the objects, it 
is simply ignored. A possible solution might be the effect 
of reacceleration, e. g. by hydromagnetic turbulences due 
to interactions of the jet material with the surrounding 
medium. Fermi acceleration by hydromagnetic turbulences 
will only affect the lowest-energy particles of the pair 
ensemble simulated here because for higher particle energies 
radiative losses strongly dominate reacceleration. Since 
the optical regime is close to the low-frequency turnover 
of the synchrotron component, reacceleration will thus 
have its dominant effect in this spectral range.

\section{Discussion of alternative models}

Until now, we have focused on the homogeneous, leptonic jet
model characterized by instantaneous injection events of 
relativistic plasma blobs into the jet. We are now discussing
a variety of alternative models which have also met with
considerable success in explaining broadband spectra of
blazars. For models invoking continuous injection of electrons
into the jet, such as the accelerating-jet model discussed
by Georganopoulos \& Marscher (\markcite{georg98}1998), the
final conclusions would be very similar to those found in
the analysis of our model since our fits require jet
filling factors close to unity.

\subsection{Shock-in-jet models}

In the previous section we had already briefly mentioned that
a shock propagating through the jet (e. g., \cite{mg85}, 
\cite{kirk98}) would be strongly SSC dominated, and 
the broadband spectrum predicted by this model would 
be very similar to our SSC dominated case (see Fig.
\ref{ssc_fit}). However, the spectrum could look drastically
different if an external radiation component becomes efficient
at large distances from the central engine. This external radiation
component could most plausibly be provided by a cloud, reflecting
part of the synchrotron radiation produced within the jet (the
synchrotron mirror, \cite{gm96,bd98}). B\"ottcher \& Dermer
(\markcite{bd98}1998) have shown that the energy density of the
reflected synchrotron radiation in the comoving frame of the
disturbance may be approximated as

\begin{equation}
{u'_{Rsy} \over u'_{sy}} \approx 4 \, \Gamma^3 \, \tau_{T, c}
\, {R'_B \over \Delta r_c} \, \left( 1 - {2 \, \Gamma \, R'_B 
\over z} \right), 
\label{u_rsy}
\end{equation}
where $\Delta r_c$ and $\tau_{T, c}$ are the radial extent and
the Thomson depth of the reflecting cloud of cold material, and
$z$ is the distance of the disturbance from the central engine.
In this model, $R'_B$ may be identified with the extent of 
the disturbance along the jet (denoted $x$ in \cite{mg85}).
Approximating further $u'_{sy} \approx \tau_{T, d} \langle \gamma
\rangle^2 \, u'_B$, where $\tau_{T, d}$ is the Thomson depth of
the disturbance along the jet axis, and $\langle\gamma\rangle^2
\, \Gamma^2 \approx \epsilon_{Rsy} / \epsilon_{sy}$, we find

\begin{equation}
{L_{Rsy} \over L_{sy}} \approx 4 \, \tau_{T, d} \, \tau_{T, c}
\, \Gamma \, {R'_B \over \Delta r_c} \, {\epsilon_{Rsy} \over
\epsilon_{sy}} \, \left( 1 - {2 \, \Gamma \, R'_B \over z}
\right).
\label{L_Rsy}
\end{equation}

Eq. \ref{L_Rsy} indicates that the synchrotron mirror becomes 
efficient, if $z \gg \Gamma \, R'_B$. In the following, we will
require this condition to be fulfilled. The variability time 
scale of a high-energy $\gamma$-ray flare in this scenario is 
given by $\Delta t_{var} = \max\left\lbrace \Delta r_c / 
(\Gamma^2 c) \, , \> R'_B / (\Gamma c) \right\rbrace$, which 
yields a constraint on $\Delta r_c$ in addition to the constraint
on $R'_B$ as given in Eq. \ref{RB}. Using these constraints, we 
find

\begin{equation}
\tau_{T, d} \, \tau_{T, c} \gtrsim {1 \over 4} \, {L_{RSy} \over L_{sy}}
\, {\epsilon_{sy} \over \epsilon_{RSy}} \sim 8 \cdot 10^{-12},
\label{tau_estimate}
\end{equation}
which is appears to be a reasonable number. Assuming that the hard
X-ray / soft $\gamma$-ray spectrum of BL~Lacertae is dominated by
the SSC process --- as we expect in the shock-in-jet model ---, and
that the SSC component peaks at $2 \lesssim \epsilon_{SSC} \lesssim
200$, we derive a bulk Lorentz factor of $10 \lesssim \Gamma \lesssim 
100$. Then, the condition $z \gg c \, \Gamma^2 \, \Delta t_{var}$ 
yields $z \gtrsim 1$~pc. Since the disturbance is emitting synchrotron
and SSC radiation basically independent of the external (reflected)
radiation field, we would expect an extended high state of

\begin{equation}
\Delta t_{sy, SSC} \sim {z \over \Gamma^2 \, c} \gg \Delta t_{var, EGRET}
\end{equation}
at mm, infrared, optical, UV, X-rays, and soft $\gamma$-rays prior
to the EGRET flare. This seems to contradict the observed 
quasi-simultaneous flaring at optical frequencies and in the EGRET 
energy range as reported by Bloom et al. (\markcite{blo97}1997), 
unless the synchrotron mirror operates at very low efficiency due 
to $z \sim 2 \, \Gamma \, R'_B$.

\subsection{Leptonic cascade models}

Blandford \& Levinson (\markcite{bl95}1995) have proposed a model
invoking a pair cascade developing along the jet as synchrotron and
inverse-Compton radiation produced by the accretion disk and within 
the jet is rescattered into the jet by circumnuclear material and
initiates a pair cascade via $\gamma\gamma$ pair production with
the inverse-Compton emission produced by relativistic electrons
injected into the jet at the injection radius. In its original
form, this model predicts single power-law high-energy spectra,
but it is well conceivable that the soft and hard $\gamma$-ray
spectra are composed of different radiation components, as in our
model. One problem with this pair cascade model is that it predicts
that the high-energy emission in the EGRET regime lags variations
at lower frequencies, and the high-energy variability should be
characterized by longer variability time scales, which is in 
contradiction to most $\gamma$-ray flares observed from blazars 
so far. This is because the $\gamma$-ray photosphere, beyond which
the jet becomes optically thin to $\gamma\gamma$ pair production
for photons of a given $\gamma$-ray energy, increases with
increasing photon energy. Also, the observed quasi-simultaneity 
of the optical and $\gamma$-ray flares of BL~Lac with probably 
roughly equal time scales might be hard to reproduce with this 
model.

A different model, also based on a leptonic pair cascade, has been
suggested by Marcowith et al. (\markcite{marcowith95}1995). In this
model, the jet consists of a mildly relativistic, cylindrical outer
electron-proton jet which powers an ultrarelativistic, inner pair
jet via hydromagnetic interaction at the boundary between these two
regions. A pair cascade is initiated by $\gamma\gamma$ absorption
of inverse-Compton scattered direct accretion disk photons with 
themselves and the radiation field of the accretion disk. The model 
predicts a strong spectral break ($\alpha_{\gamma} \approx 2 \,
\alpha_X$) at the photon energy where $\gamma\gamma$ absorption 
becomes important. As in the Blandford \& Levinson
(\markcite{bl95}1995) model, this scenario predicts a 
$\gamma$-ray photosphere increasing with $\gamma$-ray 
energy. The Marcowith et al. (\markcite{marcowith95}1995)
model has been very successful in explaining the high-energy
spectrum of 3C~273, where many other models have problems with
the large spectral break around $\sim 10$~MeV. However, in its
published form, the model does not provide detailed predictions
about the synchrotron component, and it predicts a smooth single 
power-law at high energies, inconsistent with the MeV -- GeV 
spectrum of BL~Lacertae. An additional soft photon source for
inverse-Compton scattering could possibly solve the latter
problem, but these issues would need to be addressed in detail
by the authors of this model, before a final conclusion about 
its applicability to BL~Lacertae can be drawn.

\subsection{Hadronic models}

In this subsection, we discuss models in which most of the
relativistic particle energy in the jet is primarily contained
in protons. Their energy can either be transferred to secondary
pairs through $p\gamma$ initiated pair cascades or via collisionless 
acceleration mechanisms to primary electrons.
 
Mannheim et al. (\markcite{mannheim96}1996) have presented a
model fit of the proton-initiated cascade (PIC) model of
Mannheim (\markcite{mannheim93}1993) to a non-simultaneous
broadband spectrum of BL~Lacertae. This model seems to have
the potential to provide a reasonable fit also to our simultaneous
broadband spectrum. The hardening of the $\gamma$-ray spectrum
at several GeV may be interpreted as the blue-shifted bump 
resulting from the decay of first-generation $\pi^0$ mesons 
produced in $p \gamma$ interactions. Flares at $\gamma$-ray 
energies are expected to be simultaneous to synchrotron flares, 
in agreement with the observations of BL~Lacertae. For model
parameters derived in the context of this model see Mannheim
et al. (\markcite{mannheim96}1996).

The second class of initially proton-dominated jet models is
based on plasmoid deceleration and collisionless transfer of 
internal energy of swept-up protons to non-thermal electrons
(\cite{ps99}). The radiation signatures of these processes 
have been investigated in detail by Dermer \& Chiang 
(\markcite{dc98}1998) and Chiang \& Dermer (\markcite{cd99}1999), 
and their application to blazars has been discussed in Dermer 
(\markcite{dermer99}1999) and Pohl \& Schlickeiser (\markcite{ps99}1999). 
For moderate ($\Gamma \sim 10$) bulk Lorentz factors of the plasmoid 
SSC is so far assumed to be the primary $\gamma$-ray production 
mechanism. The simultaneity between the synchrotron and the
$\gamma$-ray flare observed in BL~Lacertae suggests that the
radiative cooling time scale of non-thermal electrons in the 
blast wave is shorter than the dynamical time scale of the system,
i. e. the electrons are in the fast-cooling regime. In this case,
and if SSC scattering occurs in the Thomson regime, the $\gamma$-ray
luminosity may be approximated as

\begin{equation}
{L_{SSC} \over L_{sy}} \approx {1 \over 3} \, \left( {p - 2 \over
p - 1} \right) \, {\epsilon_e \over \epsilon_B},
\label{blastwave_ssc}
\end{equation}
where $p$ is the spectral index of the nonthermal distribution into
which electrons are accelerated, and $\epsilon_e$ and $\epsilon_B$
are the energies in nonthermal electrons and in the magnetic field
as a fraction of their respective equipartition values. For a standard
value of $p \sim 2.5$, this yields $\epsilon_e \sim 30 \, \epsilon_B$,
which implies that the magnetic field strength has to be substantially 
below the equipartition value. This has also been found for $\gamma$-ray
bursts by Chiang \& Dermer (\markcite{cd99}1999), where too strong
a magnetic field would lead to too rapid electron cooling, producing
too soft X-ray spectra, inconsistent with the hard low-energy slopes
observed by BATSE. A pure SSC spectrum, however, would not produce
the peculiar spectral shape of the high-energy spectrum observed in 
BL~Lac. A solution to this problem could be a blue-shifted $\pi^0$
bump (\cite{ps99}) or an additional contribution from Comptonization
of an external radiation field. 

\section{Summary and conclusions}

We have analyzed the multiwavelength data for BL~Lacertae during
its July 1997 outburst in terms of several popular models for
the broadband emission of blazars. We found that a reasonable
fit is only possible with a combined SSC/ECC model. The parameters 
of the central accretion disc and its environment which yield an 
acceptable fit are $M_{BH} = 2 \cdot 10^6 \, M_{\odot}$, $L_D = 
10^{44}$~erg~s$^{-1}$, $\tau_{T,BLR} = 0.025$. The minimum 
variability time scale consistent with our model is $\Delta 
t_{var, min} \sim 1.2$~h.

The need for an external inverse-Compton component in the 
GeV photon energy range, in addition to the SSC component, 
confirms that BL~Lac is an atypical BL~Lac~object. For example, 
the simultaneous broadband spectra of the X-ray selected 
BL~Lac~objects like Mrk~421 (\cite{mk97}) and Mrk~501 
(\cite{pian98}, \cite{petry99}) can very well be fitted 
with a pure SSC model, while the emergence of an 
external inverse-Compton component seems to be more 
typical of quasars, where strong line emission indicates 
the existence of relatively dense broad-line regions. In 
fact, most simultaneous broadband spectra of quasars during 
outbursts can well be fit with an ECC or ECD model (e. g., 
3C273: \cite{dss97}), or a combined SSC/ECD/ECC model 
(e. g., PKS~0528+134: \cite{bc98}, \cite{muk99}; 3C279:
\cite{hartman99}). This trend has been more firmly
established by a detailed modeling analysis of 51 
$\gamma$-ray loud blazars by Ghisellini et al. 
(\markcite{ghi98}1998).

Since BL~Lac belongs to the subclass of radio selected
BL~Lac objects whose properties are generally intermediate
between quasars and X-ray selected BL~Lacs, our results might 
strengthen the hypothesis that it is the density and structure
of matter surrounding the central engine of an AGN which 
determines its appearance as either quasar or BL~Lac
object. In particular, this implies that the differences
between these classes of objects are not predominantly due
to viewing angle effects.

Our conclusions are certainly biased in the sense that we
have concentrated on leptonic jet models of impulsive particle
injection, which is the only type of models which we have
supported by detailed numerical simulations. We have done so
because of the considerable success with which this model has met 
in explaining a wide variety of blazar spectra in many different 
$\gamma$-ray intensity states. We have discussed several 
alternative models, such as shock-in-jet models, leptonic 
cascade models and hadronic jet models. Several of these 
possibilities could not be ruled out. Detailed numerical
calculations of the application of those models to our
simultaneous broadband spectrum of BL~Lacertae are beyond 
the scope of this paper, but we strongly encourage the 
authors of those models to do similar analyses so that
a definitive conclusion in favor of a single model can 
be reached.

\acknowledgments
This research has made use of data from the University of Michigan
Astronomy Observatory, which is supported by the National Science
Foundation and by funds from the University of Michigan.
We also note that the Green Bank Interferometer is a facility
of the National Science Foundation operated by the National
Radio Astronomy Observatory in support of the NASA High Energy
Astrophysics programs.

SDB has conducted this research as a Natonal Research Council
Resident Associate. The work of MB is supported by NASA through
Chandra Postdoctoral Fellowship grant PF~9-10007 awarded by the
Chandra X-ray Center, which is operated by the Smithsonian
Astrophysical Observatory for NASA under contract NAS~8-39073.

We thank the referee for inspiring comments and suggestions,
and J. Mattox, A. Marscher and P. Sreekumar for many useful
discussions. 

\appendix

\section{Reprocessed accretion disc radiation}

For the effects of Compton scattering of accretion-disc
radiation rescattered into the jet trajectory by surrounding
clouds, the central accretion disc may be treated as a point
source, radiating isotropically, since in the process of 
rescattering in the BLR, virtually all spatial information about 
the point of origin of photons in the accretion disc gets lost.
Furthermore, the accretion disc spectrum is approximated
by a single-temperature blackbody spectrum. The blackbody
temperature $\Theta_D = kT_d / (m_e c^2)$ is determined 
in a way that the resulting spectrum peaks at the same 
photon energy as the total photon spectrum of the 
Shakura-Sunyaev (\markcite{ss73}1973) disc. This yields 
the photon production rate of the disc,

\begin{equation}
\dot N_D^{\ast} (\edstar, \odstar) = {K \over 4\pi} {{\edstar}^2 \over
e^{\edstar / \Theta_D} - 1}
\end{equation}
where $K$ is determined by the normalization to the total disc
luminosity $L_D$,

\begin{equation} 
K = {L_D \over m_e c^2} {15 \over \Theta_D^4 \, \pi^4} \> {\rm sr}^{-1}
= 1.88 \cdot 10^{49} \, {L_{44} \over \Theta_D^4} \> {\rm s}^{-1} \,
{\rm sr}^{-1}.
\end{equation}
Throughout the appendix, the asterisk denotes quantities in the
accretion disc rest frame, whereas quantities without asterisk
denote quantities in the blob rest frame. $\epsilon = h\nu /
(m_e c^2)$ is the dimensionless photon energy, and
$\odstar = (\mudstar, \phi_D^{\ast})$ is the solid angle of
emission of an accretion disc photon with respect to the
jet axis.

Following the treatment of rescattering in surrounding material
as outlined in B\"ottcher \& Dermer (\markcite{bd95}1995), but
for a time-independent situation, we find for the differential 
photon number of rescattered accretion disc photons at height 
$z$ above the disc, in the stationary frame, 

\begin{equation}
n_{ph}^{\ast} (\estar, \ostar; z) = {\sigma_T \, K \over 16 \, 
\pi^2 \, c} \, {{\estar}^2 \over e^{\estar / \Theta_D} - 1} 
\int\limits_{r_{min}}^{r_{max}} d r \> {n_e ({\bf r}) \, 
x \over r \, \vert z^2 - x^2 - r z \, \mudstar \vert},
\end{equation}
where $r$ is the distance of the scattering location in the BLR
from the center of the accretion disc (the black hole), and $x =
\sqrt{r^2 + z^2 - 2 r z \mudstar}$. $\mudstar$ follows from simple 
geometrical considerations, $n_e ({\bf r})$ is the electron density 
in the BLR at the location {\bf r} which is determined by $r$ and 
$\mudstar$. This photon density is transformed to the blob rest 
frame using the Lorentz invariance of $n_{ph} (\epsilon, \Omega) 
/ \epsilon^2 = n_{ph}^{\ast} (\epsilon^{\ast}, \Omega^{\ast}) / 
\epsilon^{\ast 2}$ 
and 

\begin{equation}
\estar = \epsilon \, \Gamma \, (1 + \bg \mu),
\end{equation}
\begin{equation}
\mustar = {\mu + \bg \over 1 + \bg \mu},
\end{equation}

where $\mu$ and $\mustar$ are the angle cosine under which the
photon enters the blob in the comiving and the stationary frame,
respectively. The photon spectrum (A3) is used in the procedure 
described in BMS to calculate the electron energy-loss rate and 
the emitted photon spectrum due to inverse-Compton scattering. 

If the blob is located well within the inner boundary of the 
BLR, $z \ll r_{in,BLR}$, the energy-loss rate can be well 
approximated assuming that all rescattered photons enter the 
blob from the front, i. e. with $\mu = -1$. This is the head-on 
approximation. Furthermore, in this case we approximate the 
photon spectrum by a delta-function in photon energy, 

\begin{equation}
{\epsilon^3 \over e^{{\epsilon \over \Theta_D} \Gamma (1 + \bg\mu)} 
- 1} \longrightarrow B_2 \> \delta(\epsilon - \langle\epsilon\rangle)
\end{equation}
where
\begin{equation}
\langle\epsilon\rangle \approx 2.7 \, {\Theta_D \over 
\Gamma (1 + \bg\mu)}
\end{equation}
and
\begin{equation}
B_2 = {\pi^4 \over 15} \, \left( {\Theta_D \over \Gamma (1 + \bg\mu)} 
\right)^4.
\end{equation}

Combining these two approximations yields
$$
- \left( {d\gamma \over dt} \right)_{ECC}^{\delta, head-on} =
\gamma^2 \> {\pi^2 \over 960} \, \sigma_T^2 \, K \, \Theta_D^4
\, \Gamma^2 \, (3 + \bg^2) \> \cdot
$$
\begin{equation}
I(\langle\epsilon\rangle), \gamma, -1) \, \int\limits_{r_{min}}^{r_{max}}
dr \, {x \, n_e ({\bf r}) \over r \, \vert z^2 - x^2 - r z \mudstar
\vert},  
\end{equation}
where $\langle\epsilon\rangle = 2.7 \Gamma \Theta_D$. The integral
$I$ involves the angle integrations over the Klein-Nishina cross
section, to which an analytical solution is given in BMS. In our 
simulations, this approximation is used as long as $z < 0.2 \cdot 
r_{in, BLR}$. In this range, its deviation from the exact electron 
energy-loss rate is negligible for all electron energies.

If

\begin{equation}
\gamma_2 \ll {1 \over 2.7 \, \Gamma \, \Theta_D},
\end{equation}
the entire scattered photon spectrum may be calculated in the Thomson
regime. In this case, it is given in the blob rest frame by

\begin{equation}
\dot n_{ECC}^{Th} (\es, \os) = {\sigma_T^2 \, K \over 32 \, \pi^2}
\int\limits_{-1}^1 d\mu \int\limits_1^{\infty} d\gamma \>
{n(\gamma) \over \gamma^2} {\epsilon^2 \over e^{\epsilon{\Gamma
(1 + \bg \mu) \over \Theta_D}} - 1} \> \int\limits_{r_{min}}^{\infty}
dr \> {n_e ({\bf r}) \, x \over r \, \vert z^2 - x^2 - r \, z \, \mudstar
\vert }
\end{equation}
where $\epsilon = \es/(\gamma^2 [1 - \beta\chi])$ and 
$\chi = \mu\mu_s$ is the cosine of the collision angle between
the scattering electron and the photon before scattering. Here,
we have neglected the very weak dependence of the scattered
photon spectrum on the azimuthal angles involved. In our
simulations, we use Eq. (A11) if $\gamma_2 \le 0.1/(2.7 \, \Gamma
\, \Theta_D)$.

\newpage

\newpage

\newpage

\end{document}